\RequirePackage{fix-cm}
\documentclass[smallcondensed]{svjour3}       
\smartqed  

\usepackage[numbers]{natbib}
\usepackage[utf8]{inputenc} 
\usepackage[T1]{fontenc}    
\usepackage{hyperref}       
\usepackage{url}            
\usepackage{booktabs}       
\usepackage{amsfonts}       
\usepackage{nicefrac}       
\usepackage{microtype}      
\usepackage{lipsum}
\usepackage{amsmath}
\usepackage{mathrsfs}
\usepackage{cases}
\usepackage[linesnumbered,ruled]{algorithm2e} 
\usepackage{indentfirst}
\usepackage{graphicx}
\usepackage{caption}
\usepackage{multirow}
\usepackage{threeparttable}
\usepackage{comment}
\usepackage{float}

\usepackage[T1]{fontenc}
\usepackage{lmodern}

\newtheorem{myDef}{Definition}
\usepackage{geometry}
\begin{document}

\title{Bidirectional group random walk based network embedding for asymmetric proximity
}


\author{Jiawei Shen  \and
        Xincheng Shu \and Hu Yang* \thanks{*Corresponding author.} 
}


\institute{Jiawei Shen
            \at College of Mathematics and Statistics, Chongqing University, Chongqing, 401331, China,
            \\\email{jwshen@cqu.edu.cn}
           \and 
           Xincheng Shu
           \at Institute of Cyberspace Security, Zhejiang University of Technology, Zhejiang, 310023, China,
           \\\email{sxc.shuxincheng@foxmail.com}
           \and 
           Hu Yang
           \at College of Mathematics and Statistics, Chongqing University, Chongqing, 401331, China,
            \\\email{yh@cqu.edu.cn}
}

\date{Received: date / Accepted: date}

\maketitle

\begin{abstract}
Network embedding aims to represent a network into a low dimensional space where the network structural information and inherent properties are maximumly preserved. Random walk based network embedding methods such as DeepWalk and node2vec have shown outstanding performance in the aspect of preserving the network topological structure. However, these approaches either predict the distribution of a node's neighbors in both direction together, which makes them unable to capture any asymmetric relationship in a network; or preserve asymmetric relationship in only one direction and hence lose the one in another direction. To address these limitations, we propose bidirectional group random walk based network embedding method (BiGRW), which treats the distributions of a node's neighbors in the forward and backward direction in random walks as two different asymmetric network structural information. The basic idea of BiGRW is to learn a representation for each node that is useful to predict its distribution of neighbors in the forward and backward direction separately. Apart from that, a novel random walk sampling strategy is proposed with a parameter $\alpha$ to flexibly control the trade-off between breadth-first sampling (BFS) and depth-first sampling (DFS). To learn representations from node attributes, we design an attributed version of BiGRW (BiGRW-AT). Experimental results on several benchmark datasets demonstrate that the proposed methods significantly outperform the state-of-the-art plain and attributed network embedding methods on tasks of node classification and clustering. 
\keywords{Network embedding \and Random walk \and Network structure \and Node attribute}
\end{abstract}

\section{Introduction} 

Networks are ubiquitous in various real-world scenarios, e.g., social networks, citation networks, biology networks.
Mining useful information in networks can benefit lots of real-word applications, such as node classification\cite{ji2011ranking,li2020hierarchical}, node clustering\cite{sun2009ranking} and link prediction\cite{gao2011temporal}, and thus has drawn wide attention in both academia and industry. Network embedding provides an effective and efficient way to mine information in networks. Generally, it aims to represent a network into a low dimensional vector space where the topological structure and inherent properties of the network can be preserved as much as possible. 

One of the most common challenges of network embedding comes from how to define the proximity between two vertexes. The first-order and second-order proximities are two common similarities in the network embedding literature\cite{cai2018comprehensive,goyal2018graph}. The first-order proximity indicates that nodes connected by each other should be similar and hence  can be considered as a proximity to preserve local network structure. However, the observed links often only cover a small portion on real-world networks. Hence, the first-order proximity is not sufficient to preserve the network structure. The second-order proximity is an important complement, which assumes that nodes with similar distribution of neighbors should also have similar representations.  
How to preserve the pre-defined proximity is another challenge in network embedding. Some earlier work like LLE\cite{belkin2001laplacian} and LE\cite{roweis2000nonlinear} preserve the first-order proximity by factorizing graph Laplacian eigenmaps. However, they suffer from both computational and statistical drawbacks. Another set of studies (\cite{ahmed2013distributed,cao2015grarep,ou2016asymmetric}) try to directly factorize the node proximity matrix.
But these methods are still not scalable for real-world networks due to high time complexity.

\begin{figure*}
\centering
\noindent\includegraphics[width=0.9\textwidth]{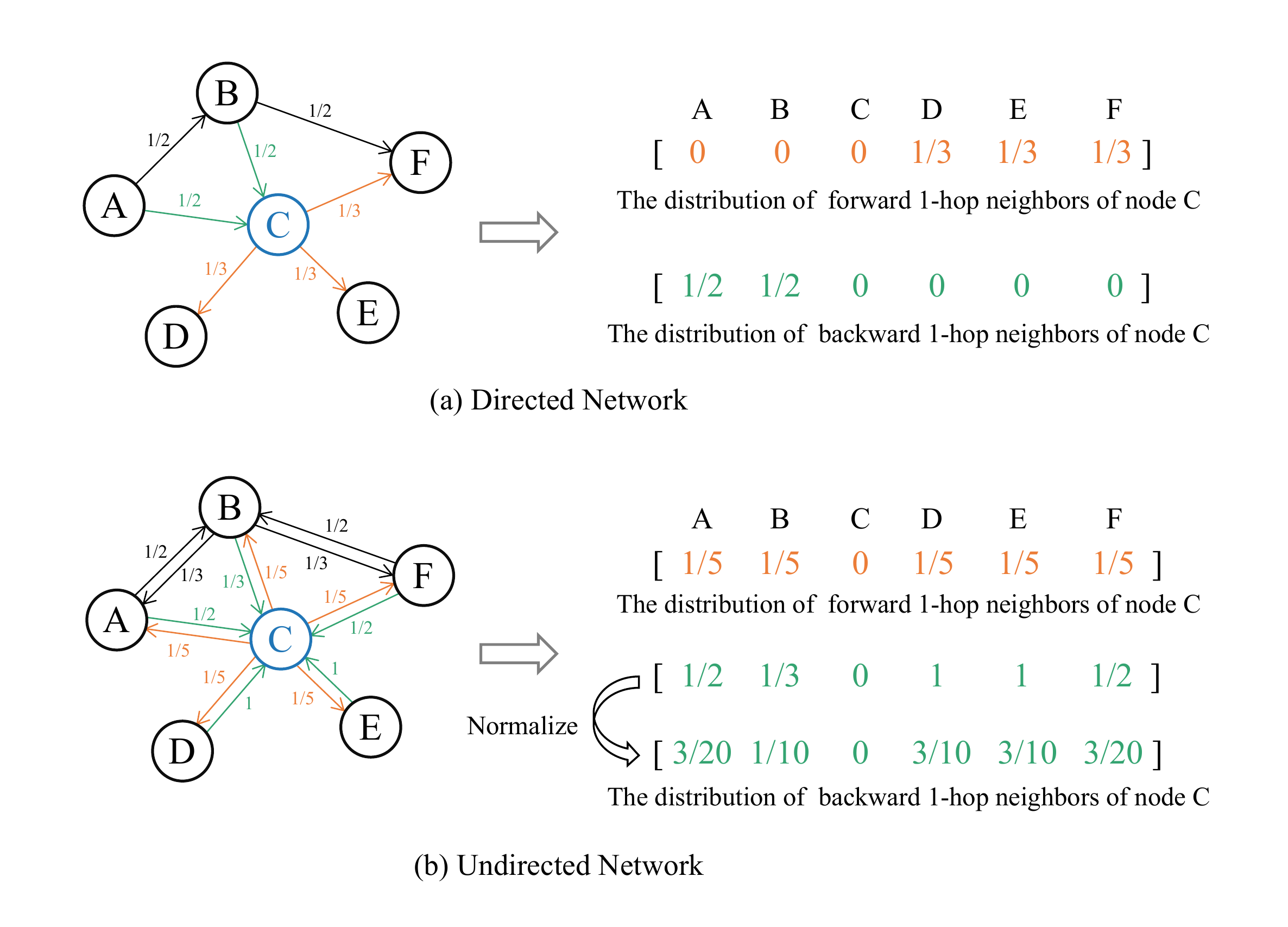}
\caption{The distributions of neighbors in the forward and backward directions for directed and undirected networks.} \label{fig:BiGRW}
\end{figure*}

Random walk based network embedding methods such as DeepWalk \cite{perozzi2014deepwalk} and node2vec \cite{grover2016node2vec} have shown outstanding  performance in the field of preserving network topological structure. In general, they adopt random walk sampling to generate neighbors for nodes in a network, which allow them to explore deeper vertices, then maximize the probability of observing the sampled neighbors for nodes conditioned on representations using SkipGram\cite{mikolov2013efficient} model. In this way, the second-order proximity is implicitly preserved. 
Both of DeepWalk and Node2vec have an assumption that a source vertex and target vertex have a symmetric effect over each other. However, this assumption often doesn't hold true in real world applications. For example, in social network like Twitter, supposing that user $A$ follows user $B$, the importance of $B$ for $A$ can't simply equal to that of $A$ for $B$. Obviously, those who a user follows and his followers are two kinds of different asymmetric relational information. Generally, We observe that neighbors of a node can be divided into two categories: neighbors in the forward direction of random walks or neighbors in the backward direction. The former refers to vertices that can be accessed in random walks starting from that vertex, while the later refers to vertices from which can access that vertex in random walks. For directed networks, the structure of neighbors of a node in the forward and backward direction is usually different.
For undirected networks, though a node have same neighbors in the forward and backward direction, they differ in the importance of each neighbor. Hence, in both directed and undirected networks, the distributions of these two kinds of neighbors are different and represent two kinds of asymmetric structural information. Figure~\ref{fig:BiGRW} illustrates the difference between the distributions of forward and backward 1-hop neighbors in both directed and undirected networks. It can be seen that the probability of $A$ arriving at $C$ does not equal to that of $C$ arriving at $A$ due to their asymmetric local structure. However, both of DeepWalk and node2vec don't distinguish a node's neighbors by the forward and backward direction. Instead, they predict the distribution of neighbors in both direction together for each node conditioned on its embedding, which makes the occurrence number of vertex pair $(A, C)$ and $(C, A)$ be same. Hence, they can not capture any asymmetric relationship in a network.

Several researchers focus on preserving various asymmetric relationships in a network. For example,  
APP\cite{zhou2017scalable} employs random walks on a network with restart and treats vertex pairs only in the forward direction as positive pairs, which allow it able to preserve the asymmetric relationship in the forward direction. GraRep\cite{cao2015grarep} preserves the asymmetric relationship in the backward direction under matrix factorization framework. However, these approaches only consider asymmetric relationship in either forward or backward direction, and hence would lose the asymmetric relationship in another direction.
Finding that the distributions of a node's neighbors in the forward and backward direction are two kinds of asymmetric structural information, we propose learning a representation for each vertex to predict the distributions of its neighbors in the forward and backward direction separately.
To our best knowledge, this is the first attempt to preserve asymmetric structural information in both forward and backward direction of random walks.  

Observing that neighbors with different distances may have different significance, we propose weighted average of $1$-$k$ steps transitivity  proximity ($k$-WAT) and give its mathematical definition in section~\ref{sec:APPROACH}. Due to the high-efficiency of random walks in terms of time and space complexity\cite{grover2016node2vec}, we design a group random walk sampling strategy to implicitly preserve the proposed $k$-WAT proximity. Specifically, we divide neighbors into different groups according to the hops between two nodes. For example, given two random walk sequences: $u_1 \rightarrow u_4 \rightarrow u_3 \rightarrow u_2$ and $u_1 \rightarrow u_2$, $u_2$ is a 3-hop and 1-hop neighbor of  $u_1$ in the former and later sequence respectively. At each random walk, we randomly choose the group of $l$-hop $(l\leq k)$ with the probability proportional to $\alpha^l$. The variation coefficient $\alpha$ actually controls a trade-off between BFS and DFS. Section~\ref{sec:APPROACH} details the group random walk sampling strategy. Intuitively, small $\alpha$ would inspire walks to search towards nodes that are highly interconnected with the starting node and hence can benefit capturing community structure.
On the contrary, large $\alpha$ would encourage walks to explore neighbors with longer distances and can help capture structural equivalence. Structural equivalence refers to that nodes having similar structural roles such as bridges or hubs in a network should have similar representations. Real-world networks often involve both of community structure and 
structural equivalence, though, their impact may vary with specific applications. Our method can flexibly control the trade-off between preserving community structure and structural equivalence based on specific cases.

Attributed network embedding has attracted wide attention in recent years. The idea is to learn low dimensional node representations that preserve both the network topological structure and attributes. The network attributes often contain some valuable information and can be helpful for some applications.  Various approaches (e.g., TADW\cite{yang2015network}, GCN\cite{kipf2016semi}, GraphSAGE\cite{hamilton2017inductive}) have been proposed for attributed networks. However, since attribute information is usually associated with specific tasks,  representations that preserve node attributes can not be generalized across different applications and the improvements of attributed network embedding approaches highly rely on the quality and relevance of attributes. In addition, the attribute information is often missing in real-world networks, which makes attributed network embedding methods hard to generalize. 
Despite these limitations, attributed network embedding approaches might have better performance for applications where high-quality and relevant attributes are available. Therefore, we design an attributed version of BiGRW (BiGRW-AT) to learn representations from node attributes. The basic idea is to obtain the target vector for each node from its attributes. The experimental results demonstrate that our method is superior to the existing attributed network embedding methods on both node classification and clustering.

To summarize, the main contributions of this paper are as follow:
\begin{itemize}
    \item We propose a novel unsupervised network embedding model called BiGRW, which can effectively capture asymmetric structural information in both forward and backward direction of random walks. We also design an attributed version of BiGRW (BiGRW-AT) to learn representations from attribute information.
    \item We propose a novel proximity called $k$-WAT and design a group random walk algorithm via SkipGram model to implicitly preserve the proposed proximity with a parameter $\alpha$ to control the trade-off between BFS
    and DFS. 
    \item We conduct experiments on several benchmark datasets. Experimental results demonstrate that our methods significantly outperform the existing unsupervised network embedding methods on the tasks of node classification and clustering.
    
\end{itemize}

The rest of this paper is organized as follow. In Section \ref{sec:Related work}, we briefly introduce related work on network embedding. In Section \ref{sec:APPROACH}, we present a mathematical definition of $k$-WAT proximity and detail the algorithm of our method. We present the experimental setting and results in Section \ref{sec:EXPERIMENTS} and conclude the paper in Section \ref{sec:CONCLUSIONS}. 

\section{Related works}
\label{sec:Related work}
\subsection{Plain network embedding}
Plain network embedding focuses on learning node representations by preserving the network topological structure. Various approaches have been proposed. These approaches can be roughly divided into matrix factorization based methods and deep learning based methods.

Matrix factorization based network embedding methods aim to learn node embedding by factorizing the network property matrix(e.g., node proximity matrix). For example, \citet{ahmed2013distributed} propose decomposing the adjacency matrix of a network to preserve the first order proximity. 
HOPE\cite{ou2016asymmetric} uses generalized Singular Value Decomposition(SVD) to preserve various higher order proximities such as Katz Index and Rooted Page Rank.
GraRep\cite{cao2015grarep} factorizes the $k$-step probability transition matrix by SVD to obtain embedding for different $k$ separately and concatenates them together after training.  

The methods based on the deep learning framework can be further divided into two categories according to whether random walk is adopted to sample node sequences from a network. The basic idea of random walk based approaches is to generate a node's neighbors by random walk sampling and then maximize the probability of observing the neighbors of a node conditioned on its representation. For example,
DeepWalk\cite{perozzi2014deepwalk} firstly applies truncated random walks to sample node sequences and learns node representations from the sequences by using SkipGram model. 
Node2vec\cite{grover2016node2vec} extends DeepWalk by proposing a more flexible sampling strategy with two parameters to control the trade-off between BFS and DFS. 
The second class of deep learning based network embedding methods directly adopt deep learning framework  to preserve network topological structure  without random walks. For example, 
LINE\cite{tang2015line} treats immediately connected vertices as neighbours when defining the second order proximity. It preserves the first and second order proximities separately and concatenates them after training. 
SDNE\cite{wang2016structural} preserves the second order proximity by reconstructing the adjacency matrix under the encoder-decoder framework, and in the meantime minimize the difference between hidden representations of connected nodes. Other examples in this category include \cite{cao2016deep,chen2018harp,li2016discriminative}.

\subsection{Attributed network embedding}
Attributed network embedding aims to learn node representations from both the network topological structure and node attributes.
A wide variety of attributed network embedding approaches have been proposed in recent years. Generally, they can be roughly divided into three categories. The first category is based on the matrix factorization framework. For example, TADW\cite{yang2015network} employs matrix factorization to combine the network topological structure and node attributes. AANE\cite{huang2017accelerated} decomposes the attribute affinity matrix with a penalty of embedding difference between connected nodes. Other examples in this category contain \cite{bandyopadhyay2019outlier,pan2020flgai,yang2019low}.

The second category applies auto-encoder framework. For example, DANE\cite{gao2018deep} uses two auto-encoders to preserve the network topological structure and node attributes respectively with a modality strategy to combine them. ANRL\cite{zhang2018anrl} employs an auto-encoder with node attributes as input to reconstruct the attributes and in the meanwhile predicts the neighborhood. Other examples using the auto-encoder framework include \cite{fan2021attributed,li2021deep,liu2018content,liu2019network}.

The third category is built upon graph convolutional networks, where node attributes are propagated along network edges. For example, GCN\cite{kipf2016semi} employs the graph convolution operation among the immediate neighbors of each node to obtain the semi-supervised network embedding. GraphSAGE\cite{hamilton2017inductive} extends GCN with several aggregators to aggregate node attributes from a sampled set of immediate neighbors. GAE and VGAE\cite{kipf2016variational} use GCN to learn node representations that are helpful to reconstruct the network adjacency matrix. Other examples in this category include \cite{chiang2019cluster,pei2020geom,velivckovic2017graph}

\section{The proposed approach}
\label{sec:APPROACH}
In this section, we present the proposed approach. Table~\ref{tab:notations} lists the main symbols used in this paper.

\begin{table}[htbp]
\centering
\caption{Terms and Notations}
\label{tab:notations}
\begin{tabular}{lll}
\\[-2mm]
\hline
\hline\\[-2mm]
\vspace{1mm}
{\bf \small Symbol}&{\bf\small Meaning}\\[+1mm]
\hline
\\[-2mm]
\vspace{1mm}
$V$      &   the set of vertices\\
\vspace{1mm}
$W^{(k)}$      &   the $k$-WAT matrix\\
 \vspace{1mm}
$W_{ij}^{(k)}$      & the element at i-th row and j-th column of $W^{(k)}$ \\
 \vspace{1mm}
$\Tilde{W}_{i\cdot}^{(k)}$, $\Tilde{W}_{\cdot i}^{(k)}$  &   the normalized i-th row and i-th column of $W^{(k)}$ respectively \\
 \vspace{1mm}
$\Tilde{W}_{i\cdot}^{(k)}(j)$, $\Tilde{W}_{\cdot i}^{(k)}(j)$    &	the j-th element of $\Tilde{W}_{i\cdot}^{(k)}$ and $\Tilde{W}_{\cdot i}^{(k)}$ respectively\\
 \vspace{1mm}
$S_i$ & the embedding source vector for vertex $v_i$ \\
 \vspace{1mm}
$T_i^{(f)}$, $T_i^{(b)}$ & the forward and backward target vector for vertex $v_i$ respectively \\
\vspace{1mm}
$X_i$ & the features of vertex $v_i$  \\
\vspace{1mm}
$H_f$,$H_b$ & the forward and backward weight matrix respectively \\
\hline
\hline
\end{tabular}
\end{table}

\subsection{Model overview}
\begin{figure*}
\noindent\includegraphics[width=\textwidth]{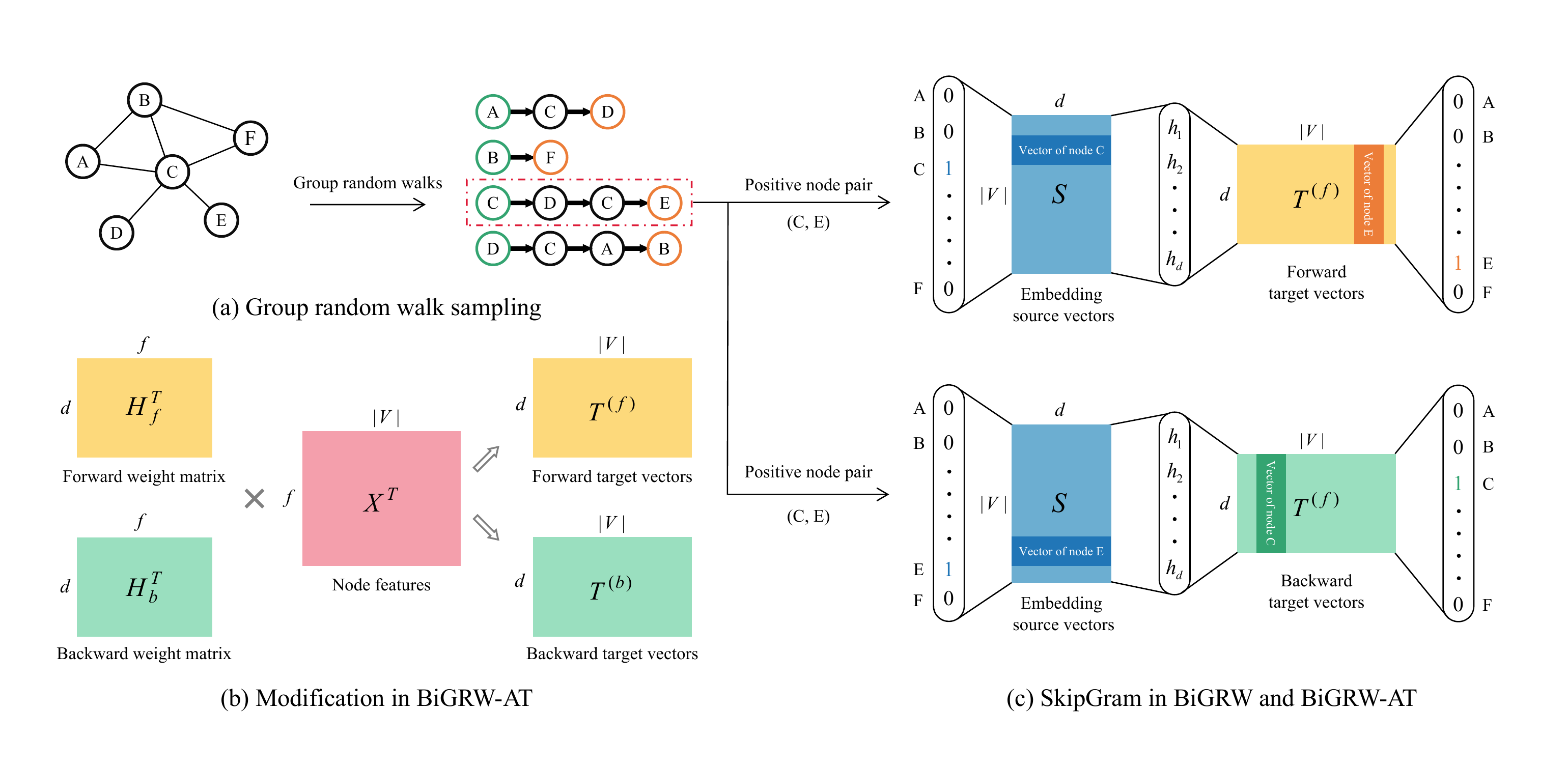}
\centering
\caption{The overview of BiGRW and BiGRW-AT. For positive node pair $(C,E)$, the starting node $C$ and the ending node $E$ in a random walk sentence are treated as the source node and target node  respectively in the forward direction, which is opposite in the backward direction. Both share the same embedding source vectors $S$. For BiGRW-AT,  the target vectors  for each direction is modified as the product of the node feature matrix and a trainable weight matrix. \label{fig:overview}}
\end{figure*}

Figure~\ref{fig:overview} shows the overview of the proposed BiGRW and BiGRW-AT. Specifically, we first simulate group random walks on the network to obtain positive node pairs. The starting node and ending node at each walk sequence are treated as a positive node node. For per node pair, the starting node and ending node are considered as the source node and target node respectively in the forward direction. The situation is opposite in the backward direction.  We then pass the positive node pairs to the SkipGram model to update the representations of the corresponding nodes. Negative sampling is adopted at this stage. For BiGRW-AT, the target vectors for each direction is modified as the product of the node feature matrix and a trainable weight matrix.

\subsection{The forward and backward $k$-WAT proximities}

The second-order proximity assumes that vertices with similar distributions of neighbors are similar to each other. We find that a node's neighbors can be divided into two categories: neighbors in the forward direction of random walks or neighbors in the backward direction. The former refers to vertices that can be accessed from that vertex in random walks, while the later refers to vertices from which can access that vertex. Apart from that, neighbors with different distances may have different significance. Based on these two observations, we define the forward and backward $k$-WAT distributions of a node's neighbors in Definition~\ref{def 1}. 

\begin{myDef}
\label{def 1}
($k$-WAT matrix)
Let $A$ be the probability transition matrix in a random walk, $A_{ij}$ is the transition probability from vertex $i$ to vertex $j$ with one step. We observe that $A^k$ is the k-step probability transition matrix and $A_{ij}^k$ is exactly the transition probability from vertex $i$ to vertex $j$ after $k$ steps. The $k$-WAT matrix is defined as follow:
\begin{equation*}
    W^{(k)} = \alpha_1A^1+\alpha_2A^2+\cdots+\alpha_kA^k
\end{equation*}
where $\alpha_k$ is the weight for the group of k-hop , s.t., $ \alpha_1+\alpha_2+\cdots+\alpha_k=1$.
Specially, $\alpha_l=\frac{\alpha^l}{\sum_{i=1}^k \alpha^i}(l \leq k)$ in BiGRW, where $\alpha$ is the variation coefficient.

\end{myDef}

For vertex $v_i$, the forward $k$-WAT distribution of its neighbors in the forward direction is defined as the normalized $i$-th row of $k$-WAT matrix $W^{(k)}$; while the $k$-WAT distribution of its neighbors in the backward direction is defined as the normalized $i$-th column of $W^{(k)}$. Naturally, the forward and backward $k$-WAT proximities can be defined as follow:

\begin{myDef}
\label{def 2}
(Forward k-WAT proximity)
The forward $k$-WAT proximity between vertex $v_i$ and $v_j$ is defined as the similarity of $\Tilde{W}_{i\cdot}^{(k)}$ and $\Tilde{W}_{j\cdot}^{(k)}$, where $\Tilde{W}_{i\cdot}^{(k)}$ and $\Tilde{W}_{j\cdot}^{(k)}$ are the the normalized i-th row and j-th row of $W^{(k)}$ respectively.
\end{myDef}

\begin{myDef}
\label{def 3}
(Backward k-WAT proximity)
The backward $k$-WAT proximity of $v_i$ and $v_j$ is defined as the similarity of $\Tilde{W}_{\cdot i}^{(k)}$ and $\Tilde{W}_{\cdot j}^{(k)}$, where $\Tilde{W}_{\cdot i}^{(k)}$ and $\Tilde{W}_{\cdot j}^{(k)}$ are the normalized i-th column and j-th column of $W^{(k)}$ respectively.
\end{myDef}

Obviously, the forward and backward $k$-WAT proximities are two kinds of asymmetric structural relationships. Thus we try to learn a representation for each node that is useful to predict the $k$-WAT distribution of its neighbors in the forward and backward direction separately.

\subsection{Loss functions}

Formally, let $p_i(\cdot)=(p_i(1),p_i(2),\cdots,p_i(|V|))$ denote the pre-defined $k$-WAT distribution of neighbors in the forward or backward direction for vertex $v_i$, $q_i(\cdot)=(q_i(1),q_i(2),\\ \cdots,q_i(|V|))$ denote the predicted distribution conditioned on the representation of vertex $v_i$.
Note that $p_i(\cdot)$ is exactly $\Tilde{W}_{i\cdot}^{(k)}$ for the forward direction, or $\Tilde{W}_{\cdot i}^{(k)}$ for the backward direction. The term $q_i(j)$ for the forward direction is defined as:
\begin{equation*}
    q_i(j)=\frac{\exp(S_i\cdot T_j^{(f)})}{\sum_{k \in V}\exp(S_i\cdot T_k^{(f)})},
\end{equation*}
while $q_i(j)$ for the backward direction is defined as:
\begin{equation*}
    q_i(j)=\frac{\exp(S_i\cdot T_j^{(b)})}{\sum_{k \in V}\exp(S_i\cdot T_k^{(b)})},
\end{equation*}
where $S_i$ is the embedding source vector for vertex $v_i$, while $T_i^{(f)},T_i^{(b)}$ are the forward and  backward target vectors for vertex $v_i$, respectively.

To preserve the corresponding proximity, we minimize the distance between $p_i(\cdot)$ and $q_i(\cdot)$ :
\begin{equation*}
    \mathcal{L}=\sum_{i \in V}d(p_i(\cdot),q_i(\cdot)),
\end{equation*}
where $d(\cdot,\cdot)$ is the distance measure between two distribution. Adopting KL-divergence as the distance measure and omitting some constants, we have :
\begin{equation*}
    \mathcal{L}_{f}=-\sum_i \sum_j \Tilde{W}_{i\cdot}^{(k)}(j)\log q_i(j), 
\end{equation*}
\begin{equation*}
    \mathcal{L}_{b}=-\sum_i \sum_j \Tilde{W}_{\cdot i}^{(k)}(j)\log q_i(j),
\end{equation*}
where $\mathcal{L}_{f}$ and $\mathcal{L}_{b}$ are the objective functions for the forward direction and backward direction respectively. 

\subsection{Optimization via group random walk sampling}
Due to the high-efficiency of random walks in terms of both time and space complexity, we propose a group random walk sampling strategy to implicitly optimize the above objective, where the probability of visiting vertex $v_j$ starting from vertex $v_i$ within walk steps $k$ is exactly the term $W_{ij}^{(k)}$. 
Algorithm~\ref{algo:algorithm} details the group random walk sampling process. For each walk, we randomly choose the walk length $l(l\leq k)$ with the probability proportional to $\alpha^l$ and then simulate a walk with steps $l$ to get a starting vertex $u$ and an ending vertex $v$. Note that $u$ and $v$ are treated as the source vertex and the target vertex respectively in the forward direction, which is opposite in the backward direction, i.e., $v$ is the source vertex and $u$ is the target vertex. 

Applying group random walk sampling, the forward $k$-WAT proximity in a network can be preserved in the latent representation space by maximizing the probability of observing the neighbors in the forward direction. The backward $k$-WAT proximity can be preserved likewise. The objective functions for the forward and backward directions are modified as following:
\begin{equation*}
    \mathcal{L}_{f}=-\sum_{(i,j)\in \mathcal{S}} \log q_i(j),
\end{equation*}
\begin{equation*}
    \mathcal{L}_{b}=-\sum_{(i,j)\in \mathcal{S}}\log q_j(i),
\end{equation*}
where $\mathcal{S}$ is the sampling set of positive node pair, and $(i,j)\in \mathcal{S}$ indicates that a walk path starting from vertex $v_i$ and ending at vertex $v_j$ being sampled in random walks.
Since summation over all inner product with each vertex in a network is computationally expensive, we adopt negative sampling method to improve training efficiency, which distinguishes the target nearby vertices for each vertex from other vertices using logistic regression.
Using negative sampling, we obtain the final objective functions:
\begin{equation}
\label{forward objective}
    \mathcal{L}_{f}=-\sum_{(i,j)\in \mathcal{S}}\Big( \log \sigma(S_i \cdot T_j^{(f)})
    +\sum_{k=1}^QE_{v_k\sim P_n}[\log\sigma(-S_i \cdot T_{v_k}^{(f)})]\Big),
\end{equation}

\begin{equation}
\label{backward objective}
    \mathcal{L}_{b}=-\sum_{(i,j)\in \mathcal{S}}\Big( \log \sigma(S_i \cdot T_j^{(b)})\\
    +\sum_{k=1}^QE_{v_k\sim P_n}[\log\sigma(-S_i \cdot T_{v_k}^{(b)})]\Big),
\end{equation}
where $Q$ is the number of negative nodes that are sampled for each positive node pair. $P_n$ is the noise distribution for drawing negative samples. 

\begin{algorithm}[htbp]
	\caption{Training Algorithm for BiGRW}
	\label{algo:algorithm}
	\KwIn{$G(V,E,W)$, variation coefficient $\alpha$, largest walk steps $k$, number of iterations $I$, number of negative samples $Q$ }
	\KwOut{the embedding source vectors $S$, the forward target vectors $T^{(f)}$, and the backward target vectors $T^{(b)}$}
	Initialize:$S$, $T^{(f)}$, $T^{(b)}$\;
	
	\For{epoch=1 to I}{	
	    $\mathcal{O}$ = Shuffle($V$)\;
		\For{each $u \in \mathcal{O}$}
		{
		    
		    $l$ = SampleWalkLength($k$, $\alpha$)\;
		    $v$ = SampleEndPoint($G,u,l$)\;	$L_f=-\log \sigma(S_u \cdot T_v^{(f)})$\;
		    $L_b=-\log \sigma(S_v \cdot T_u^{(b)})$\;
		    $Loss=L_f+L_b$\;
		    \For{j=1 to Q}{
		$neg$=RandomUniform($V$)\;
		$L_f=-\log \sigma(-S_u \cdot T_{neg}^{(f)})$\;
		    $L_b=-\log \sigma(-S_v \cdot T_{neg}^{(b)})$\;
		    $Loss+=L_f+L_b$\;
		}
		UpdateByAdam($Loss$,  $(S, T^{(f)}, T^{(b)})$)\;
		}
}
\end{algorithm}

\subsection{Attributed version of BiGRW}
In order to capture the node attributes, we design an attributed version of BiGRW. The objective function remains the same as formula~\ref{forward objective} and \ref{backward objective}. The only difference lies in the target vectors $T\in \mathcal{R}^{|V| \times d}$, which is modified as the product of the node feature matrix $X \in \mathcal{R}^{|V|\times f}$ and a trainable weight matrix $H\in \mathcal{R}^{f \times d}$, i.e., $T=XH$. For the forward and backward target vector of vertex $v_i$, $T_i^{(f)}=X_i H_f$, $T_i^{(b)}=X_i H_b$, respectively.

\subsection{Complex analysis}
The time complexity of the proposed BiGRW model framework consists of the random walk sampling process and the optimization via SkipGram model.

We can generate a positive node pair by simulating a random walk of length $l\leq k$. Hence, the time complexity of random walk process is $\mathcal{O}(kI|V|)$, where $k$, $I$, $|V|$ is the largest walk steps, the number of iterations and the number of nodes, respectively. For each node pair, the complexity of calculating loss objective function is $\mathcal{O}(d^2)$ for each direction, where $d$ is the embedding dimension. Therefore, the complexity of the SkipGram part in BiGRW is $\mathcal{O}(d^2IQ|V|)$, where $Q$ is the number of negative samples. For BiGRW-AT, the operation of calculating the target vector from node attributes requires extra $\mathcal{O}(df^2)$ per vertex, where $f$ is the dimension of attributes.

In summary, the overall complexity of BiGRW and BiGRW-AT is $\mathcal{O}(kI|V|+d^2IQ|V|)$ and $\mathcal{O}(kI|V|+d^2IQ|V|+df^2IQ|V|)$, respectively. By ignoring the parameters $d$, $k$, $f$, $I$ and $Q$ which are often small constants, the complexity of the proposed BiGRW and BiGRW-AT can be equivalent to $\mathcal{O}(|V|)$, which is scaling linearly w.r.t. the number of nodes in the network. Hence, our proposed model is computationally efficient and can be well applied on large-scale networks with millions of nodes and above.

\section{Experiments}
\label{sec:EXPERIMENTS}

In this paper, we conduct experiments on three benchmark networks and evaluate the performance of various network embedding methods on tasks of node classification, node clustering and visualization. The experimental results demonstrate that our proposed methods performs better than the baseline methods.

\subsection{Experimental setup}
\subsubsection{Datasets}
In order to evaluate the effectiveness of network embedding methods, we collect three benchmark datasets, including two citation networks and one social network\footnote{https://linqs.soe.ucsc.edu/data}. Table~\ref{tab:networks} summarizes the description about these networks.
\begin{itemize}
    \item Cora and CiteSeer\cite{yang2015network}: They are citation networks of scientific publications, where nodes indicate academic publications, and each edge represents a citation relationship between two publications. 
    \item BlogCatalog\cite{tang2009relational}: It's a social network of online users, where nodes represent bloggers on the BlogCatalog website, and each edge corresponds to a relationship between two bloggers.
\end{itemize}

\begin{table}[htbp]
 \caption{Statistics of the real-world networks}
  \centering
  \resizebox{0.6\textwidth}{!}{
  \begin{tabular}{ cccc }
    \toprule
    
    \multirow{2}*{Name} & \multicolumn{2}{c}{Citation Network}  & Social Network                \\
    \cmidrule(lr){2-4} 
        & Cora     & CiteSeer 
    & BlogCatalog \\
    \midrule
    Nodes & 2708  & 3327 & 10312   \\
    Edges & 5429 & 4732 & 333983    
    \\
    Labels & 7 & 6 & 39
    \\
    Attributes & 1433 & 3703 & -
    \\
    \bottomrule
  \end{tabular}
  }
  \label{tab:networks}
\end{table}
\subsubsection{Baseline methods}
Since BiGRW mainly focuses on preserving the network topological structure, we firstly compare it with the existing plain network embedding methods that also preserve the network topological structure. These baseline approaches are presented as follow:
\begin{itemize}
    \item DeepWalk/node2vec\cite{grover2016node2vec,perozzi2014deepwalk}: These two approaches adopt random walks to sample node sequences and use SkipGram model to learn node representations from the sampled sequences.
    We use $p=1$,$q=1$ in node2vec for comparison.
    \item LINE\cite{tang2015line}: it  preserves the first-order and second-order proximities separately and concatenates them after training.
    \item GraRep\cite{cao2015grarep}: it factorizes the $k$-steps ($k=1,2,3,\cdots$) probability transition matrix by SVD separately and concatenates them together after training. 
    \item HOPE\cite{ou2016asymmetric}: it preserves various higher order proximities such as Katz Index and Rooted Page Rank by using generalized SVD.
    \item SDNE\cite{wang2016structural}: it preserves the second order proximity by reconstructing the adjacency matrix under the encoder-decoder framework, and in the meantime minimize the difference between hidden representations of connected nodes.
\end{itemize}

For scenarios where node attributes are available, we compare the attributed version of BiGRW (BiGRW-AT) with the following unsupervised attributed network embedding approaches. 
\begin{itemize}
    \item Auto-encoder: We use an auto-encoder to preserve the node attributes only as the baseline method.
    \item GAE and VGAE\cite{kipf2016variational}: These two approaches employ graph convolutional networks (GCNs) to propagate node features along edges on a network and learn representations to reconstruct the network adjacent matrix.
    \item TADW\cite{yang2015network}: it jointly preserves the network topological structure and node attributes under the framework of matrix factorization.
    \item GraphSAGE\cite{hamilton2017inductive}: it extends GCN with several aggregators to aggregate node attributes from a sampled set of immediate neighbors. We use its unsupervised version with mean-based aggregators. 
    \item DANE\cite{gao2018deep}:it uses two auto-encoder to preserve the structural and attributed information respectively, and employs a cross modality strategy to combine the topological sturcture and attributes. 
\end{itemize}

\subsubsection{Parameters and reproducibility}
All the methods are implemented in Pytorch. For all the baseline methods except Auto-encoder, we use the implementations provided by either the authors or open source projects on Github. We integrate all the implementations into an same framework\footnote{https://github.com/thunlp/OpenNE/tree/pytorch. For the proposed BiGRW and BiGRW-AT, we will make the code publicly available upon acceptance} where each model is evaluated using unified testing code after obtaining its embedding. For fairness, the final embedding dimension of all the models is set to 256 on Cora and Citeseer, 128 on BlogCatalog. For models that concatenate two vectors as the final embedding, each vector is set to 128 on Cora and Citeseer, 64 on BlogCatalog. For node2vec, the two important parameters are set as $p=1$ and $q=1$. For LINE, we concatenate the first-order and second-order representations to obtain the final embedding. For GraRep, the maximum step is set to 4.
For GraphSAGE, we remove the rectified linear units (ReLU) in the output layer since we observe that this modification can benefit its performance in experiments.
Other hyper-parameters are set as the default values provided in either the corresponding papers or open source projects.
The major hyper-parameter settings for BiGRW and BiGRW-AT are summarized in Table~\ref{tab:parameter setting}.

\begin{table}[htbp]
    \caption{Major hyper-parameters in BiGRW and BiGRW-AT.}
  \label{tab:parameter setting}
  \centering
  \resizebox{0.9\textwidth}{!}{
  \begin{threeparttable}
  
    \begin{tabular}{ ccccc }
    \toprule
    Methods&Dataset&Variation coefficient $\alpha$&Largest walk steps $k$& Embedding dimension $d$\cr
    \midrule
    \multirow{3}{*}{BiGRW}&
    Cora&1.25&5&256\cr
    &Citeseer&1.25&6&256\cr
    &BlogCatalog&0.75&5&128\cr
    \midrule
    \multirow{2}{*}{BiGRW-AT}&
    Cora&0.75&6&256\cr
    &Citeseer&0.5&6&256\cr
    \bottomrule
    \end{tabular}
    
    \end{threeparttable}
    }
\end{table}

\subsection{Experimental results}
\subsubsection{Node classification}
Node classification is one of the most common tasks to evaluate the effectiveness of node embedding in the literature. In the node classification setting where part of nodes in a network have already been attached one or more labels, the goal is to predict labels for the left nodes based on the representations. Specifically, we randomly sample 10 \% to 90 \% of the labeled nodes as training samples and the remaining nodes are used as testing samplings. We repeat this process 10 times and compute the average value of both $Micro\text{-}F_1$ and $Macro\text{-}F_1$. For all approaches, we adopt a one-vs-rest logistic regression implemented by Scikit-learn for classification.

\begin{figure*}
\noindent\includegraphics[width=\textwidth]{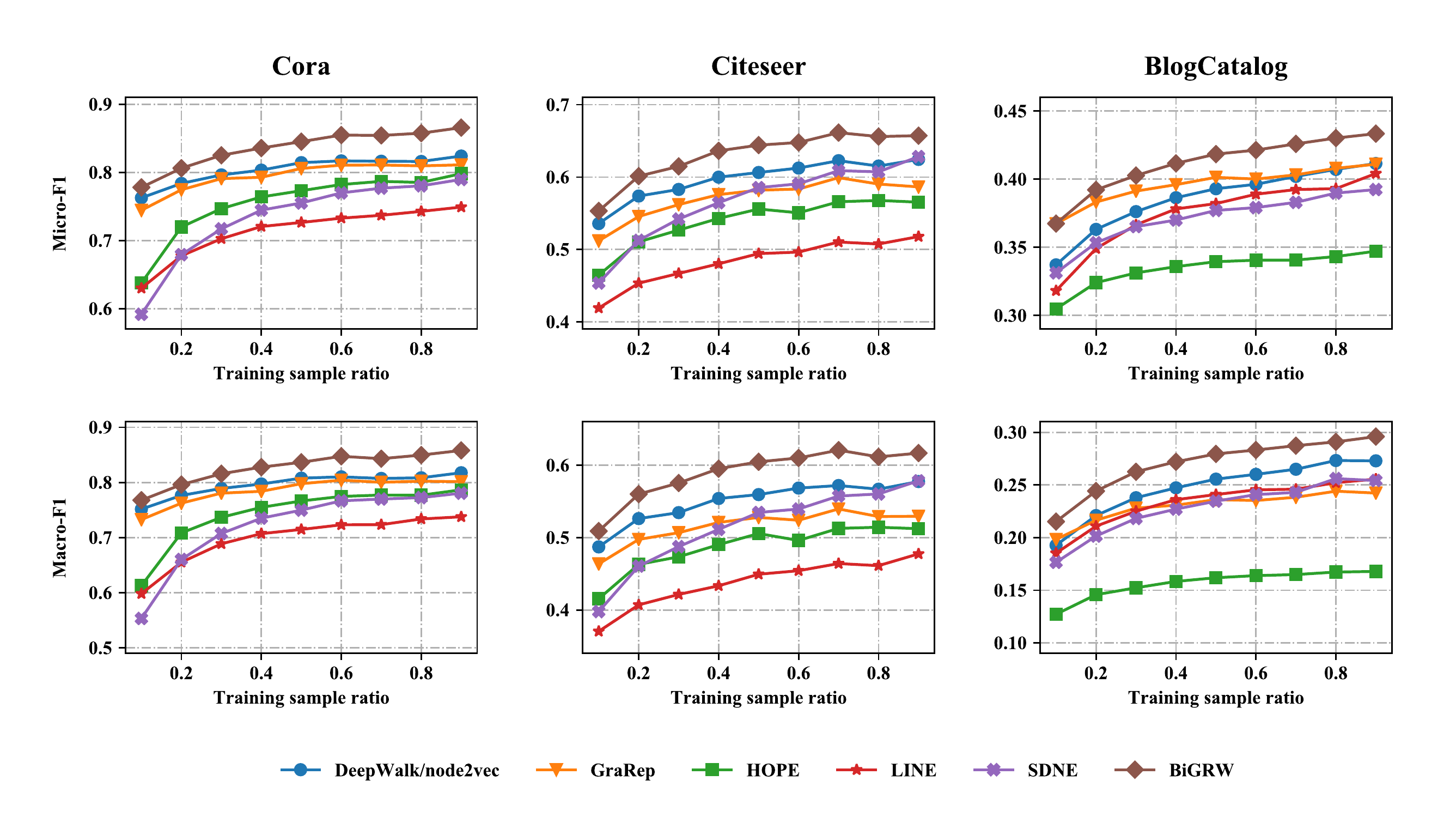}
\centering
\caption{\centering{Node classification results of plain network embedding methods.} \label{fig:plain}}
\end{figure*}

Figure~\ref{fig:plain} shows the performance of plain network embedding methods on node classification.
From the results, we can see that the proposed BiGRW has better performance than the baseline methods in all cases, which demonstrates the effectiveness of BiGRW in the aspect of preserving network topological structure. In addition, random walk based network embedding methods (DeepWalk/node2vec and BiGRW) perform relatively well on citation networks. Compared with DeepWalk/node2vec, BiGRW achieves a noticeable improvement in all cases. Our explanation is as follow. DeepWalk/node2vec predict neighbors of nodes in both direction together and hence they can't preserve any asymmetric structural information. On the contrary, BiGRW distinguishes neighbors between the forward and backward directions. It predicts the distribution of neighbors in the forward and backward direction separately, which allow it to capture asymmetric topological structure in both direction. Apart from that, the parameter $\alpha$ in BiGRW makes it flexibly control the trade-off between capturing the community structure and structural equivalence. The poor performance of LINE on citation networks is possibly because it only treat immediately connected nodes as neighbors, which only cover a small proportion of nodes and hence are not sufficient to preserve the network structure.

\begin{figure*}
\noindent\includegraphics[width=0.9\textwidth]{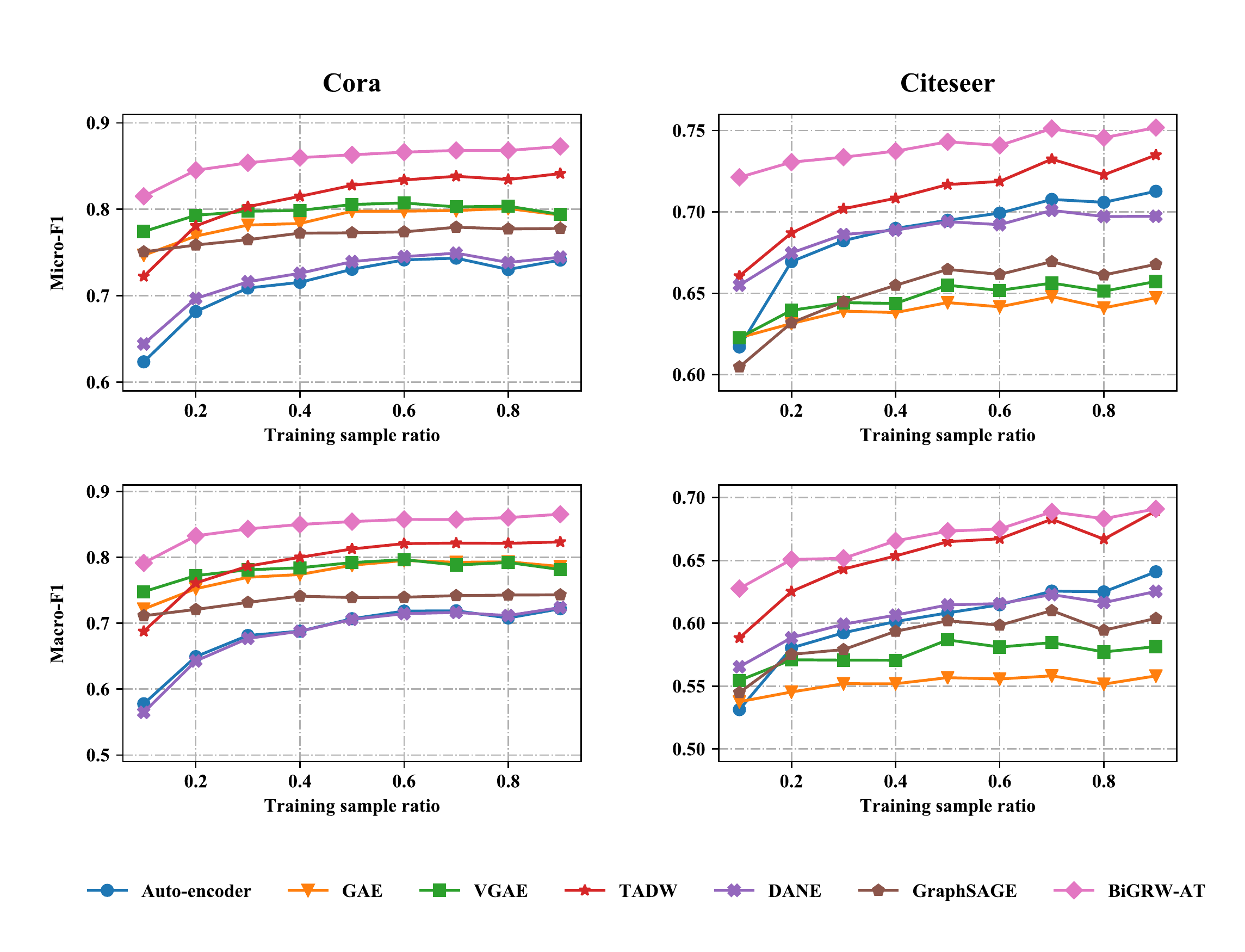}
\centering
\caption{Node classification results of unsupervised attributed network embedding methods.} 
\label{fig:attribute}
\end{figure*}

We evaluate the performance of unsupervised attributed network embedding methods in node classification on Cora and Citeseer, where node attributes are available.  Figure~\ref{fig:attribute} shows the results. As one can see, the proposed BiGRW-AT significantly outperforms the unsupervised attributed network embedding baseline approaches on both Cora and Citeseer too. It demonstrates that the proposed bidirectional mechanism and group random walk sampling strategy remain effective when network attributes are incorporated.    

\subsubsection{Node clustering}
The task of node clustering aims to group similar node together.
To further evaluate the effectiveness of network embedding, we conduct experiments on node clustering for each method after obtaining the node representations. Specifically, we apply the K-means clustering algorithm on the node embedding and evaluate the clustering results with the metrics of purity, normalized mutual information (NMI) and Matthews correlation coefficient (MCC). Table~\ref{tab:performance_structure}
shows the results of plain network embedding approaches. It can been seen that the proposed BiGRW is superior to the baseline approaches in all datasets. The results of unsupervised attributed network embedding approaches are shown in Table~\ref{tab:performance_attribute}. As we can see, the proposed BiGRW-AT significantly outperforms all the baseline approaches in all cases. For example, in terms of NMI metric, BiGRW obtains 23.5\% relative improvement compared with the best competitor (GAE) on Cora; on Citeseer, the relative improvement is 22.4\% compared with the best competitor (DANE). 

\begin{table}[htbp]

  \centering
  \begin{threeparttable}
  \caption{Node clustering results of plain network embedding methods.}
  \label{tab:performance_structure}
    \begin{tabular}{ccccccccc}
    \toprule
    \multirow{2}{*}{Methods}&
    \multicolumn{3}{c}{Cora}&\multicolumn{3}{c}{Citeseer}&\multicolumn{2}{c}{BlogCatalog}\cr
    \cmidrule(lr){2-4} \cmidrule(lr){5-7} \cmidrule(lr){8-9}
    &Purity&NMI&MCC&Purity&NMI&MCC&Purity&NMI\cr
    \midrule
    DeepWalk/node2vec&0.6237&0.4414&0.3644&0.4435&0.2104&0.1263&0.4554 &0.3476\cr
    LINE&0.5118&0.3092&0.1479&0.3312&0.0954&0.0346&0.4294&0.3303\cr
    GraRep&0.5679&0.3783&0.2771&0.3566&0.0940&0.0544&0.4445&0.3371\cr
    HOPE&0.3730&0.1406&0.0100&0.2585&0.0525&0.0305&0.3673&0.2936\cr
    SDNE&0.4306&0.2157&0.0720&0.3022&0.0817&0.0203&0.3663&0.2665\cr
    BiGRW&{\bf 0.6680}&{\bf 0.4691}&{\bf 0.4041}&{\bf 0.4828}&{\bf 0.2609}&{\bf 0.1300}&{\bf 0.4754}&{\bf 0.3566}\cr
    \bottomrule
    \end{tabular}
    \begin{tablenotes}
    \item[*] As MCC is only applicable for binary classification, we don't report the MCC result on BlogCatalog, where some of nodes have more than one labels.
    \end{tablenotes}
\end{threeparttable}
\end{table}

\begin{table}[htbp]
  \caption{Node clustering results of unsupervised attributed network embedding methods.}
  \label{tab:performance_attribute}
  \centering
  \resizebox{0.8\textwidth}{!}{
  \begin{threeparttable}
    \begin{tabular}{ccccccc}
    \toprule
    \multirow{2}{*}{Methods}&
    \multicolumn{3}{c}{Cora}&\multicolumn{3}{c}{Citeseer}\cr
    \cmidrule(lr){2-4} \cmidrule(lr){5-7} 
    &Purity&NMI&MCC&Purity&NMI&MCC\cr
    \midrule
    Auto-encoder&0.3951&0.1479&0.0992&0.4783&0.2276&0.1966\cr
    GAE&0.6518&0.4660&0.3556&0.5592&0.2617&0.2551\cr
    VGAE&0.6328&0.4054&0.3205&0.4094&0.1681&0.0959\cr
    DANE&0.6030&0.3580&0.2922&0.6513&0.3531&0.3560\cr
    GraphSAGE&0.5030&0.2742&0.1687&0.4837&0.2598&0.1555\cr
    TADW&0.6592&0.4591&0.3906&0.5580&0.2694&0.2268\cr
    BiGRW-AT&{\bf 0.7441}&{\bf 0.5759}&{\bf 0.5469}&{\bf 0.7129}&{\bf 0.4321}&{\bf 0.4409}\cr
    \bottomrule
    \end{tabular}
    \end{threeparttable}
    }
\end{table}

\subsubsection{Parameter sensitivity}
In order to evaluate the sensitivity of the three major hyper-parameters (i.e., variation coefficient $\alpha$, embedding dimension $d$ and largest walk steps $k$) in BiGRW, we conduct experiments on tasks of node classification and clustering on Cora. For variation coefficient $\alpha$, we fix $d=256$, $k=5$. For embedding dimension $d$, we fix $\alpha=1.25$, $k=5$. For largest walk steps $k$, we fix $\alpha=1.0$, $d=256$. Other hyper-parameters remain invariant. 

\begin{figure}[htbp]
\noindent\includegraphics[width=\textwidth]{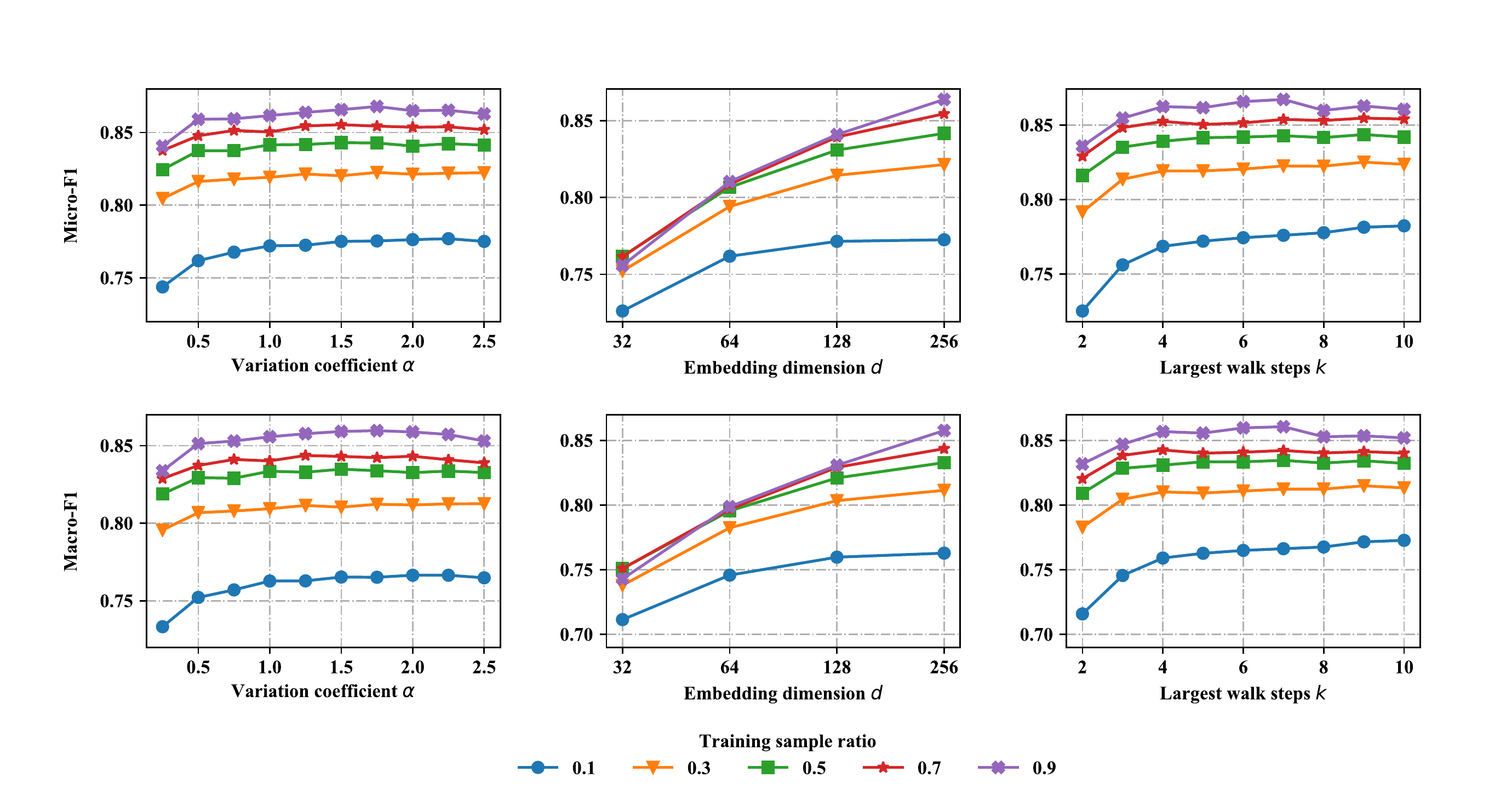}
\centering
\caption{\centering{The parameter sensitivity results in node classification on Cora.} \label{fig:sensitivity classification}}
\end{figure}

\begin{figure}[htbp]
\centering
\noindent\includegraphics[width=\textwidth]{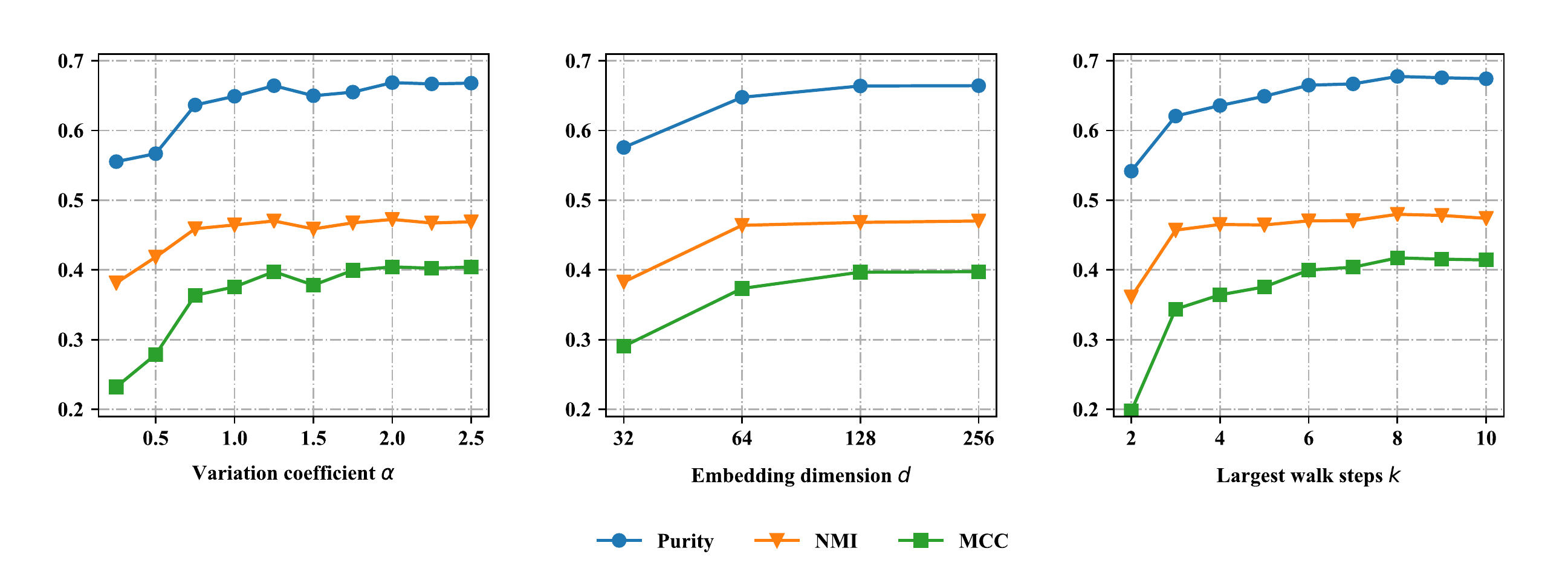}
\caption{\centering{The parameter sensitivity results in node clustering on Cora.}\label{fig:sensitivity clustering}}
\end{figure}

Figure~\ref{fig:sensitivity classification} and ~\ref{fig:sensitivity clustering} show the results in node classification and clustering, respectively.
As one can see, the performance of BiGRW on both tasks improves as $\alpha$ increases, and tends to saturate when $\alpha$ reaches around 1.25. 
The embedding dimension has relatively high impact on the performance of the method in node classification, but its influence is relatively small in node clustering. On both tasks, the trend of largest walk steps $k$ is quite similar with that of $\alpha$. Intuitively, larger $k$ allows walks to have larger search scope when defining neighbors, while $\alpha$ controls the proportion of neighbor groups of different hops, i.e., larger $\alpha$ would encourage walks to search nodes with longer distance as neighbors. Hence, larger $\alpha$ and $k$ would help BiGRW capture structure equivalence, which can be useful for node classification and clustering from the results on Cora.

\subsubsection{Visualization}

\begin{figure}[htbp]

\centering
\includegraphics[width=\textwidth]{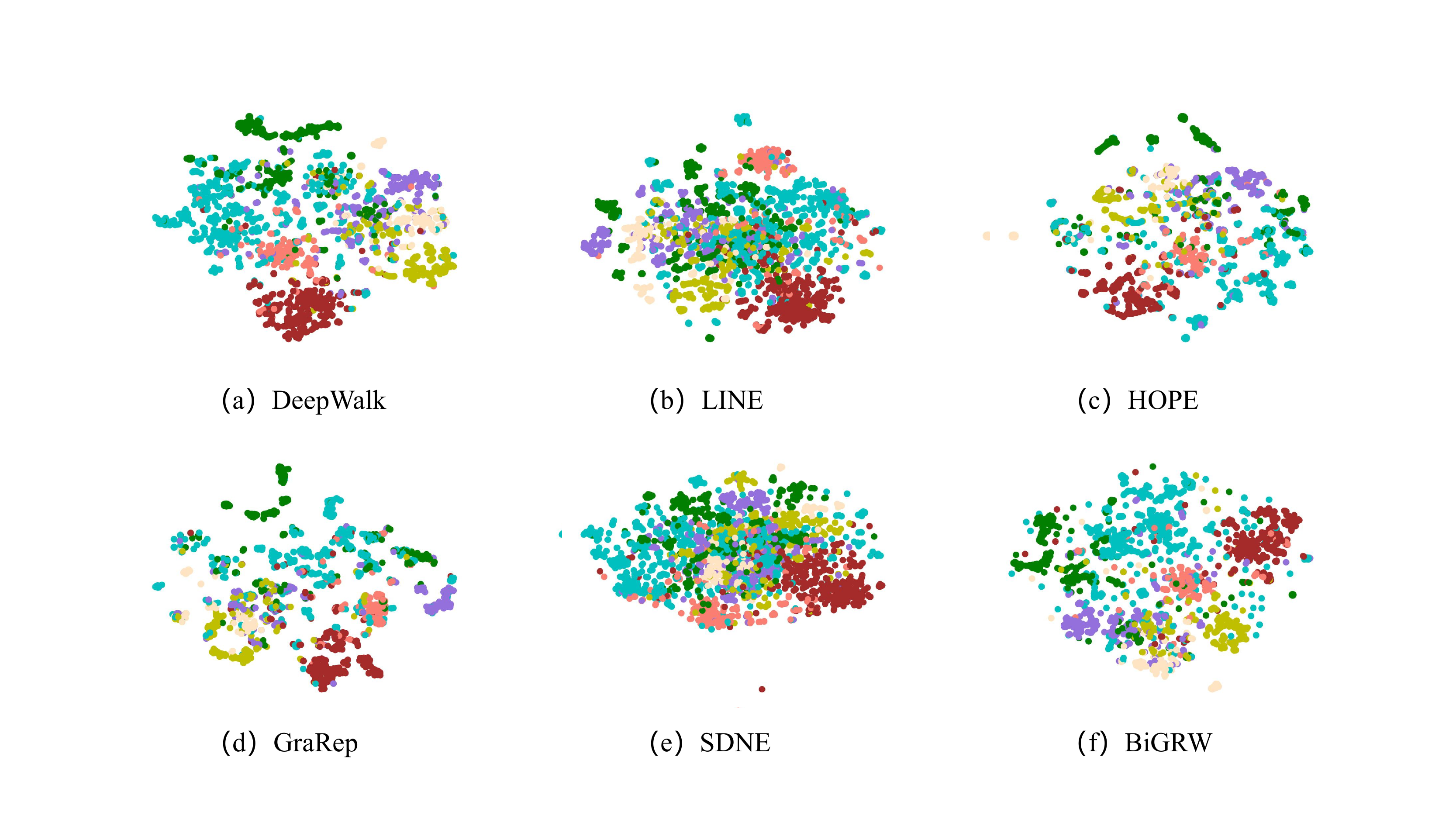}

\caption{Visualizing results of the learned node representations on Cora. Each point indicates one paper and colors of nodes represent different publication divisions.}
\label{fig:visualizing}

\end{figure}

Finally, we visualize the learned node representations of unsupervised network embedding methods that focus on preserving network topological structure by using t-SNE\cite{van2008visualizing}. Figure~\ref{fig:visualizing} shows the visualization results. BiGRW obtains relatively more compacted and separated clusters compared with the other baseline approaches, indicating the effectiveness of BiGRW in the aspect of preserving network structure again.

\section{Conclusions}
\label{sec:CONCLUSIONS}

We have noticed that the distributions of a node's neighbors in the forward and backward directions of random walks are two different asymmetric structural information. To capture asymmetric topological structure in both forward and backward directions, we proposed a novel network embedding model called BiGRW. The basic idea of BiGRW is to learn a representation for each node that is useful to predict its distribution of neighbors in the forward and backward direction separately. Apart from that, a novel random walk sampling strategy was proposed with a parameter $\alpha$ to flexibly control the trade-off between BFS and DFS. We also designed an attributed version of BiGRW to make it able to learn node representations from attributes. The experimental results demonstrate that BiGRW and its attributed version are significantly superior to the existing plain network embedding approaches and attributed network embedding approaches respectively in both node classification and clustering. 

Our future research will focus on incorporating the bidirectional mechanism and group random walk sampling strategy into more complicated networks (e.g., heterogeneous networks and dynamic networks). We may also focus on designing more complex and effective mechanism to combine the network topological structure and node attributes. 


%
%

\bibliographystyle{spbasic}      
\bibliography{refs}   

\end{document}